\documentclass[conference]{IEEEtran}
\IEEEoverridecommandlockouts
\usepackage[normalem]{ulem}
\usepackage{listings}
\usepackage{calc}
\usepackage{colortbl}
\usepackage{caption}
\usepackage{graphicx}
\usepackage{stackengine}
\usepackage{xcolor}
\usepackage{tabularx}
\usepackage{placeins}
\usepackage{multirow}
\usepackage[export]{adjustbox}
\usepackage{tabto}
\usepackage{printlen}
\usepackage{graphicx,wrapfig,lipsum}
\usepackage{soul}
\usepackage{ragged2e}
\usepackage{amsmath,amssymb,amsfonts}
\usepackage{MnSymbol,wasysym}
\usepackage{wrapfig}
\usepackage{makecell}
\usepackage{tikz}
\usepackage{pifont}

\usepackage{algorithmic}
\usepackage{graphicx}
\usepackage{textcomp}
\usepackage{xcolor}
\usepackage{hyperref}
\usepackage[a4paper, total={184mm,239mm}]{geometry}
\def\BibTeX{{\rm B\kern-.05em{\sc i\kern-.025em b}\kern-.08em
    T\kern-.1667em\lower.7ex\hbox{E}\kern-.125emX}}

\usepackage{multirow}

\definecolor{codegreen}{rgb}{0,0.6,0}
\definecolor{codegray}{rgb}{0.5,0.5,0.5}
\definecolor{codepurple}{rgb}{0.58,0,0.82}
\definecolor{backcolour}{rgb}{0.95,0.95,0.95}

\lstdefinestyle{mystyle}{
    backgroundcolor=\color{backcolour},   
    commentstyle=\color{codegreen},
    keywordstyle=\color{magenta},
    numberstyle=\color{codegray},
    stringstyle=\color{codepurple},
    basicstyle=\ttfamily\footnotesize,
    breakatwhitespace=true,         
    breaklines=true,                 
    captionpos=b,                    
    keepspaces=true,                 
    numbers=left,                    
    numbersep=4pt,                  
    showspaces=false,                
    showstringspaces=false,
    showtabs=true,                  
    tabsize=2,
    xleftmargin=2em,
    frame=lines,
    framexleftmargin=1.5em    
}

\def\qquad{\hskip2em\relax}

\usepackage{caption}
\usepackage{subcaption}
\usepackage{enumitem}
\usepackage{makecell}

\usepackage{textcomp}
\usepackage{xcolor}
\usepackage[ruled,vlined,linesnumbered]{algorithm2e}
\usepackage{algorithmic}

\usepackage{soul}
\usepackage{float}
\usepackage{tikz}
\usepackage{dblfloatfix}

\let\oldnl\nl
\newcommand\nonl{
  \renewcommand{\nl}{\let\nl\oldnl}}

\newcommand*\circled[1]{\tikz[baseline=(char.base)]{ \node[shape=circle,fill,inner sep=0.4pt] (char) {\textcolor{white}{#1}};}}

\def\BibTeX{{\rm B\kern-.05em{\sc i\kern-.025em b}\kern-.08em
    T\kern-.1667em\lower.7ex\hbox{E}\kern-.125emX}}

\begin{document}

\title{ Decoupled Access-Execute enabled DVFS for tinyML deployments on STM32 microcontrollers}

\author{\IEEEauthorblockN{Elisavet Lydia Alvanaki, Manolis Katsaragakis, Dimosthenis Masouros, Sotirios Xydis and Dimitrios Soudris}
\IEEEauthorblockA{\textit{Microprocessors and Digital Systems Laboratory, ECE, National Technical University of Athens, Greece}\\
\{ealvanaki, mkatsaragakis, dmasouros, sxydis,  dsoudris\}@microlab.ntua.gr}

\thanks{This work has been partially funded by EU Horizon program CONVOLVE under grant agreement No 101070374 (https://convolve.eu/).}}

\maketitle

\begin{abstract}
Over the last years the rapid growth Machine Learning (ML) inference applications deployed on the Edge is rapidly increasing. Recent Internet of Things (IoT) devices and microcontrollers (MCUs), become more and more mainstream in everyday activities. In this work we focus on the family of STM32 MCUs. We propose a novel methodology for CNN deployment on the STM32 family, focusing on power optimization through effective clocking exploration and configuration and decoupled access-execute convolution kernel execution. Our approach is enhanced with optimization of the power consumption through Dynamic Voltage and Frequency Scaling (DVFS) under various latency constraints, composing an NP-complete optimization problem. We compare our approach against the state-of-the-art TinyEngine inference engine, as well as TinyEngine coupled with power-saving modes of the STM32 MCUs, indicating that we can achieve up to 25.2\% less energy consumption for varying QoS levels.

\end{abstract}


\section{Introduction}
\label{sec:intro}

Over the last years, \textit{Edge AI}~\cite{li2018edge} has appeared to the forefront as a novel computing paradigm.
Edge AI refers to the deployment of complex Deep Neural Network (DNN) models directly on edge devices, such as sensors, Internet of Things (IoT) devices, and other embedded systems.
Furthermore, the increasing demand for edge-centric digital signal processing applications, such as video analytics~\cite{ananthanarayanan2017real}, mobile visual tasks~\cite{howard2017mobilenets} and others has established Convolutional Neural Networks (CNNs) as the go-to technology for efficiently handling these tasks at the edge.
Notably, major technology providers have responded to this paradigm shift by offering enterprise-grade solutions designed to optimize and deploy CNN models at the edge.
These solutions include software-based stacks (e.g., Amazon SageMaker Edge~\cite{amazonsagemakeredge}) as well as custom, purpose-built hardware chips (e.g., Google’s Edge TPU ASIC~\cite{googleedgetpu}). 

A significant amount of research efforts are focusing on exploring the complex DNN/CNN design space of dataflow schedules \cite{ZigZagMeiHJGV21, Kwon2020MAESTRO} to optimize energy. Despite their sophistication, these solutions often assume the presence of hardware accelerators or OS-supported and network-connected edge devices.
However, as the computing continuum extends to the far edge of networks, arrays of resource-constrained devices, typified by microcontrollers (MCUs), are introduced into the ecosystem.
Unlike their more capable counterparts, MCUs typically employ bare-metal application execution with custom or lightweight run-times and, at times, disconnected from the network grid.
This paradigm shift to the far edge has highlighted the concept of ``tinyML"~\cite{banbury2021micronets}.
TinyML involves the optimization of DNN/CNN models to operate seamlessly on these bare-metal MCUs, thus, extending the boundaries of where ML can be applied.
In the face of limited processing power and memory, tinyML models are designed to offer efficient inference capabilities, making it possible to bring intelligence to devices previously excluded from the ML landscape.

Deploying CNNs on MCUs introduces two major challenges.
First, the limited memory capacity of MCUs poses difficulties in both storing and executing CNN models.
This becomes even more complex, since emerging CNN architectures tend to become much deeper, aiming to provide better accuracy and/or support more complex applications~\cite{kakolyris2023road}.
Second, given that MCUs are frequently embedded in battery-operated edge devices, preserving energy resources becomes crucial, since the execution of resource-intensive and computationally hungry DNNs can rapidly deplete the battery, particularly concerning devices with extended operational requirements.

Focusing on CNN inference, prior research~\cite{chitty2022neural} and widely-used frameworks (e.g., TFLite Micro~\cite{david2021tensorflow}, Microsoft's NNI~\cite{microsoftnni}, ARM's CMSIS-NN~\cite{lai2018cmsis}, etc.) offer optimization techniques that can enable the deployment of tiny CNN models on resource-constrained MCUs.
These techniques typically include static, model-specific, post-training optimizations, such as mixed precision arithmetic ~\cite{capotondi2020cmix},  model's weight pruning~\cite{de2022depth} and quantization~\cite{moosmann2023flexible,spantidi2023automated}, as well as downsampling~\cite{saha2020rnnpool} and Neural Architecture Search (NAS) methodologies~\cite{banbury2021micronets,liberis2021munas}.
Other works examine schedule and source code optimizations either to enable efficient data and computation mapping on the underlying hardware~\cite{lin2020mcunet, jaiswal2023minun} or to divide CNN layers into independently distributable tasks that can be offloaded and executed in parallel~\cite{zhao2018deepthings}.



Although a vast amount of optimizations have been investigated for DNN deployment on MCUs, they mostly focus on models' structural features, e.g., weights, arithmetic, architecture etc., while little attention is given to system-level runtime optimization knobs.  
In the context of power consumption and/or energy efficiency optimization, Dynamic Voltage and Frequency Scaling (DVFS)~\cite{snowdon122009platform} forms an efficient tuning knob to balance performance and energy consumption by right-adjusting the clock frequency of the MCU according to the computational demands of the deployed model. 
Still, the customization of DVFS policy to DNN specific features as well as the materialization of DVFS in practice is not straightforward, as it may impose switching overheads~\cite{acun2019fine} not straightforwardly analyzable and increased leakage power due to longer execution times~\cite{jejurikar2004leakage}. 
In this work, we introduce an end-to-end methodology that exploits DVFS for optimizing the energy consumption of CNN inference on low-end MCUs. The proposed methodology is built upon three pillar concepts, i.e. i) selection of energy optimal MCU clocking scheme among differing iso-latency configurations, ii) employment of Decoupled Access Execute (DAE) computing scheme for CNN convolution layers to unlock higher DVFS efficiencies and iii) extraction of DVFS optimal allocation decisions tailored to target CNN architectures and MCU.    
More specifically, we propose a Decoupled Access Execution scheme, which splits the execution of convolution layers' kernels into memory-bound and compute-bound regions, thus, allowing the exploitation of processor idling during memory accesses. We characterize this new design space for energy-efficient DNNs and we cast/formulate a knapsack optimization problem taking into consideration the MCU's clock parameters and DAE granularities to obtain the optimal DVFS strategies given the neural network architecture and the corresponding QoS constraint.
To the best of our knowledge, \textit{this is the first work that examines the application of DVFS for CNN inference on low-end MCUs.}
We note that our methodology is orthogonal to optimization frameworks proposed in the past and can be seamlessly applied to pre-/post-training optimized models to further increase the energy efficiency at runtime (e.g., optimized models exported from TFLite Micro~\cite{david2021tensorflow} or TinyEngine~\cite{lin2020mcunet}).
We showcase the validity and efficacy of our methodology by applying it on top of three CNN inference models derived from the TinyEngine and deploying the resulting model on a STM32 MCU, one of the most popular and widely used MCUs in the embedded systems domain. 
Our experimental results show that we can achieve up to 25.2\% power optimization compared to existing state-of-the-art approaches.



\section{Clocking Scheme of STM32 Microcontrollers}
\label{sec:clocking-schemes}

\begin{figure}[t]
    \centering
    \includegraphics[width=\columnwidth]{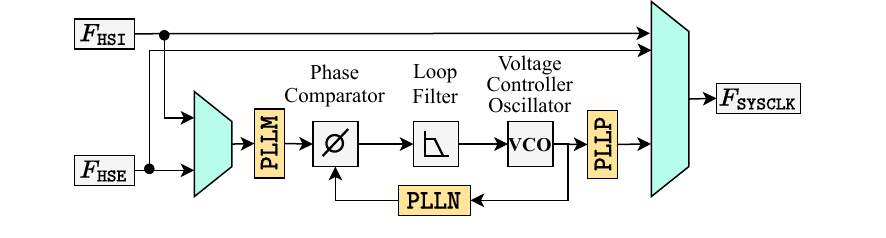}
    \caption{Simplified circuit diagram for clock configuration through \texttt{HSE} and \texttt{PLL} parameters.}
    \label{fig:stm32_clocks}
\end{figure}

The clocking system of STM32 microcontrollers is managed by the Reset and Clock Control (RCC) peripheral.
The RCC provides a wide range of clocks and clock sources which cater to various system requirements, e.g., peripheral Bus and UART clocks~\cite{gay2018beginning}.
In this work, we focus on specific clocks and settings, which determine the frequency of the system clock (\texttt{SYSCLK}) of the MCU, responsible for driving the CPU core, memory, and some peripheral modules.
Figure~\ref{fig:stm32_clocks} illustrates the simplified circuit diagram of how the \texttt{SYSCLK} is determined.
Specifically, the output frequency of the clock can be configured by the following clock sources and settings:
\begin{itemize}[leftmargin=*]
    \item \textbf{High-Speed Internal (\texttt{HSI}) Clock:} The \texttt{HSI} clock source is an internal oscillator within the MCU. The \texttt{HSI} clock operates at 16MHz by default. \texttt{SYSCLK} can be derived directly from \texttt{HSI} or through the \texttt{PLL} when \texttt{HSI} is selected as the \texttt{PLL} input source.
    \item \textbf{High-Speed External (\texttt{HSE}) Clock:} The \texttt{HSE} clock source is an external clock provided by an external crystal oscillator or clock generator. Depending on the MCU, the \texttt{HSE} clock can be configured to run at various frequencies, with our examined board supporting a range from 1 to 50MHz. Similar to the \texttt{HSI}, the \texttt{SYSCLK} can be derived directly from \texttt{HSE} or through the \texttt{PLL} when \texttt{HSE} is selected as the \texttt{PLL} input.
    \item \textbf{Phase-Locked Loop (\texttt{PLL}):} The \texttt{PLL} is a hardware module which allows to multiply the frequency of the selected input clock source (either \texttt{HSI} or \texttt{HSE}) by a programmable factor to generate a higher-frequency output. As shown in Fig.~\ref{fig:stm32_clocks}, this programmable factor is determined by input and output dividers within the \texttt{PLL} circuit, which  simplify the \texttt{PLL} design and achieve stability requirements within a wider range of input and output frequencies~\cite{gay2018beginning}. Specifically, the frequency of the system clock can be calculated by the following equation: 

    \begin{equation}
        F_{\texttt{SYSCLK}} = F_{\{\texttt{HSE,HSI}\}}*\frac{\texttt{PLLN}}{\texttt{PLLM}*\texttt{PLLP}}
        \label{eq:freq}
    \end{equation}

    where, \texttt{PLLM} is a factor that determines how much the input frequency is multiplied before it reaches the Voltage-Controlled Oscillator (\texttt{VCO}) in the \texttt{PLL}; \texttt{PLLN} is the multiplication factor for the \texttt{VCO} input frequency to determine the \texttt{VCO} output frequency, which is then provided as feedback to the Phase Comparator, aiming to provide input/output synchronization. The loop filter is utilized in order to ensure system stability and mitigate ripple effects during clock startup; Last, \texttt{PLLP} determines the division factor to obtain the final \texttt{SYSCLK} frequency.
\end{itemize}


   

\subsection{Considerations of \texttt{SYSCLK} Frequency Scaling}
\label{subsec:considerations}
Determining the frequency of the system's clock, $F_{\texttt{SYSCLK}}$, can be achieved in different ways (e.g., directly through the \texttt{HSI/HSE} clock sources or through the \texttt{PLL} circuit).
Choosing between these alternatives involves important trade-offs.
To examine the impact of each alternative on the operational efficiency and power consumption of the MCU, we develop and execute a specialized microbenchmark, designed to execute repetitive addition operations within a loop.
We focus our exploration specifically on the \texttt{HSE} and \texttt{PLL} parameters, since the \texttt{HSI} clock source yields higher power consumption compared to the \texttt{HSE} and is also prone to drift and jitter, thus, providing less stability and precision~\cite{gay2018beginning}.
We set the value of the \texttt{PLLP} equal to 2, which is the minimum possible value for the divider, since for the same $F_{\texttt{SYSCLK}}$, selecting a higher \texttt{PLLP} value leads to a higher required \texttt{VCO} frequency and, thus, higher power consumption (as evident from Fig.~\ref{fig:stm32_clocks} and Eq.~\ref{eq:freq}).

\begin{wrapfigure}{r}{4.2cm}
\vspace{-8pt}
\includegraphics[width=0.45\columnwidth]{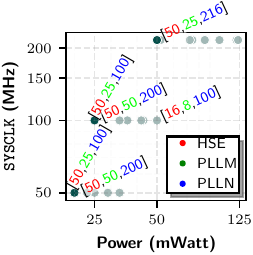}
\caption{Clock Frequency and Power for different \texttt{HSE}, \texttt{PLLM} and \texttt{PLLN} configurations.}
\label{fig:power_freq_pareto}
\vspace{-8pt}
\end{wrapfigure}

\textbf{Power consumption of iso-frequency configurations:}
Figure~\ref{fig:power_freq_pareto} illustrates the impact of different \texttt{HSE}, \texttt{PLLM} and \texttt{PLLN} configurations on the power consumption of the board for different \texttt{SYSCLK} frequencies. 
Two major observations are derived from Fig.~\ref{fig:power_freq_pareto}: (i) the same output frequency can be generated through different \texttt{HSE}, \texttt{PLLM} and \texttt{PLLN} combinations, however (ii) the selected configuration strongly affects the power consumption on the STM32 MCU. 
For instance, generating an \texttt{SYSCLK} output of 100MHz for the tuples \{$50,25,216$\} and \{$16,8,100$\} leads to 50\% power gap. The combinations that minimize
the power consumption are selected for the target \texttt{SYSCLK}. 
Different combinations may generate the same output frequency and power consumption, e.g. \{$50,25,100$\} and \{$50,50,200$\}, thus further investigation is required for selecting the optimal combination. Similar observations are derived for other \texttt{SYSCLK} configurations.

\textbf{Switching between different \texttt{SYSCLK} frequencies:} 
Generating the \texttt{SYSCLK} frequency using the \texttt{PLL} module introduces a notable switching overhead ($\approx 200\mu sec$), since, when modifying the \texttt{PLL} parameters, the circuit has to be restarted, resulting in a substantial delay per switch.
On the other hand, switching from the \texttt{PLL} frequency to the \texttt{HSE} clock occurs almost instantly, due to the direct wiring of the \texttt{HSE} with the \texttt{SYSCLK} (Fig.~\ref{fig:stm32_clocks}).
Thus, for high-to-low (i.e., $<50MHz)$) frequency switches, opting for the \texttt{HSE} clock over re-calibrating the \texttt{PLL} parameters can be beneficial.

\section{Decoupled Access Execute enabled DVFS on STM32 MCUs}
\label{sec:methodology}


\begin{figure*}[t]
    \centering
    \resizebox{0.985\textwidth}{!}{
    \includegraphics[width=\columnwidth]{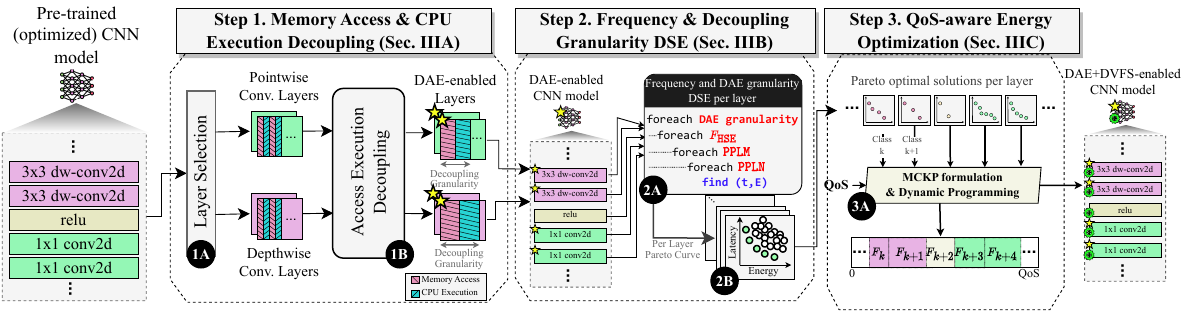}}
    \caption{Overview of the proposed methodology.}
    \label{fig:methodology}
\end{figure*}

This section introduces the proposed methodology for optimizing energy consumption of CNN model inference on a STM32 MCU under latency constraints (a.k.a. QoS).
The rationale behind our approach is that end-users often have distinct inference latency/throughput constraints for their applications and/or even operating in the context of battery-operated far-edge MCUs.
Figure~\ref{fig:methodology} shows an end-to-end overview of the proposed approach. 
The proposed methodology consists of three distinct phases (described in Sections \ref{subsec:dae} - \ref{subsec:Optimizer}), applicable to both unoptimized CNN models and optimized ones, e.g., models exported from TinyEngine~\cite{lin2020mcunet}.

\subsection{Step 1: Memory Access \& CPU Execution Decoupling}
\label{subsec:dae}
The first step of our proposed methodology focuses on source-code level restructuring, with the objective
to \textbf{D}ecouple memory \textbf{A}ccesses from CPU \textbf{E}xecution (DAE), thus, creating memory-bound and compute-bound sub-segments within the layer's structure.
DAE restructuring is a key-enabler for our strategy, as it provides more fine-grained control over when and how frequently memory accesses and computations occur (Sec.~\ref{subsec:dvfs-dse}).
This enables the application of different frequencies for extended durations and with finer control, tailored to the specific requirements of each operation, such as memory access and computation, thus, resulting in the mitigation of clock switching overhead issues.
We first identify the CNN model's most computationally-intensive and time-consuming layers (\circled{1A}).
We focus and apply DAE (\circled{1B}) on two specific layer types, i.e., \textit{i)} depthwise and \textit{ii)} pointwise convolutions.
These layer types make up over 80\% of the total number of layers found in deep lightweight CNN models, e.g., Mobilenet~\cite{howard2017mobilenets}, which employ the concept of depthwise separable convolution to reduce model's size and complexity.


\begin{lstlisting}[float,language=C, caption=Simplified source code modification for enabling decoupled access-execution (DAE) in depthwise convolutions., label=dae-snippet-dw,basicstyle=\scriptsize]
for (channel = 0; channel < in_channels/g; channel+=g) {
    //LFO for memory bound operations (Sec.IIIB)
    ClockSwitchHSE(hse);
    //Memory Bound Segment: Load g channels from the feature maps 
    q15_t buf1,...,bufg = getChannels(ch1,...,chg);
    //HFO for computation (Sec. IIIB)
    ClockSwitchPLL(pllm,plln,hse);
    // Compute Bound Segment: Perform depthwise convolution for each channel 
    convolve_depthwise(kernel, buf1, **args);
    convolve_depthwise(kernel, buf2, **args);
    ...
    convolve_depthwise(kernel, bufg, **args);
}
\end{lstlisting}


\textbf{Depthwise Convolutions}: 
Depthwise convolution is a specialized CNN operation where each input channel is convolved with a separate learnable filter, capturing spatial features per channel. 
Typically, CNNs gradually learn increasingly complex features, each one represented by a different channel. 
For instance, in an image, the initial 3 input channels (RGB) increase as the network processes the image to extract and represent more complex features, such as textures, specific objects and others.
State-of-the-art frameworks, such as CMSIS-NN~\cite{lai2018cmsis} and TinyEngine~\cite{lin2020mcunet} implement a per-channel computation approach for depthwise convolutions.
In contrast, our DAE approach introduces a parametric unrolling factor called the "decoupling granularity", denoted by $g$.
This factor determines the number of channels that are buffered in cache memory before the convolution operation is executed on each of them, thus, separating the code into memory-bound and compute-bound subsegments.
Listing~\ref{dae-snippet-dw} provides a simplified code snippet that illustrates the practical implementation of DAE optimization, showing how the decoupling granularity enables the efficient execution of depthwise convolutions.
For example, when $g=4$, four channels are fetched in the MCU's cache memory before proceeding to the actual computation.
This division enables the application of different clock frequencies per segment, which we further discuss in Sec.~\ref{subsec:dvfs-dse}.

\textbf{Pointwise Convolutions}: 
Depthwise convolution is often followed by pointwise convolution to perform channel-wise mixing and dimensionality reduction, thus, reducing model size and computational complexity while maintaining performance.
Pointwise convolutions involve \texttt{1x1} kernel sizes and are applied to each element within input channels. 
CMSIS-NN\cite{lai2018cmsis} and TinyEngine~\cite{lin2020mcunet} implement pointwise convolutions in a per-column manner. Each column consists of one element per input channel. Our approach performs decoupling on the per channel memory accesses, thus splitting the code segment into memory and compute-bound regions.
Just as in the depthwise convolution, we introduce the concept of the decoupling granularity, denoted as $g$, for modular buffering support w.r.t. the number of columns fetched from memory.
For instance, for a 8x8x3 input image and a 1x1x3 kernel, $g$ columns are loaded into the cache before performing computation for each one, in contrast with TinyEngine and CMSIS-NN, which load a single 1x1x3 image column at a time.
In general, integrating multiple buffers leads to the generation of larger memory/compute-bound regions, thus we can minimize the frequency switching overhead while avoiding high power consumption.
However, very high buffer size can lead the cache misses to skyrocket resulting in performance degradation.


After the DAE phase is performed, the CNN layers are encapsulated with a configurable decoupling granularity factor for the layer unrolling. The modified DAE-enabled CNN model is propagated to the next step of our proposed methodology (Step 2.) for effective exploration and configuration of the DAE granularity factor, alongside with the DVFS parameters exploration. DAE-enabled CNNs entail no accuracy drops.

\subsection{Step 2: DAE and Clocking Co-exploration}
\label{subsec:dvfs-dse}
At this step, we analyze the performance and energy consumption in a per-layer manner(\circled{2A}) considering the interplay effects of DAE and clocking scheme configurations.
To measure the energy consumption and performance of each layer, we have developed and integrated a custom run-time monitoring mechanism for supporting per-layer monitoring and profiling. Our mechanism relies on the on-board timers of the target MCU, which are triggered in-between the layers' code segments. Furthermore, we take advantage of STM32 MCUs integrated support for power sampling and we monitor the power consumption prior and after the DVFS integration on every CNN layer. 
The power and performance metrics for the DVFS in each layer are aggregated and utilized for the design space exploration for DAE and clocking configuration.

\begin{figure}[t]
    \centering
    \includegraphics[width=\columnwidth]{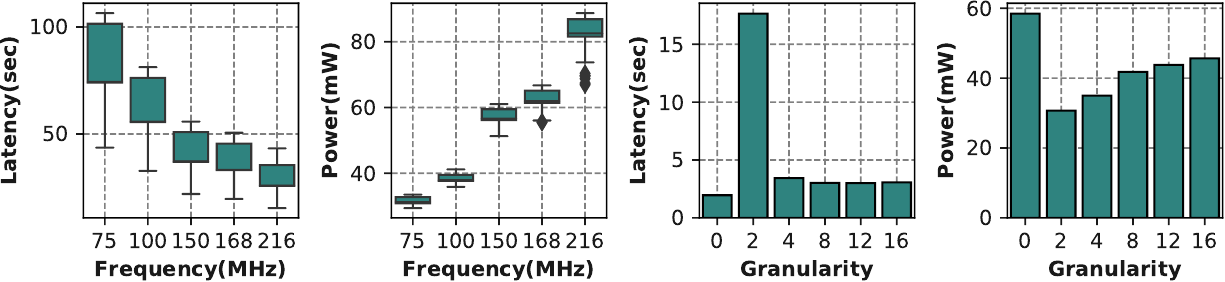}
    \caption{Impact of different DAE and clocking configurations on latency and power of depthwise and pointwise layers.}
    \label{fig:gran}
\end{figure}

More specifically, we co-explore and gain insights on the design space defined by the following three key parameters: \textit{i)} the decoupling granularity factor $g$ described in Sec.~\ref{subsec:dae}; \textit{ii)} the clock frequency of the \texttt{SYSCLK} clock; and \textit{iii)} the selection of parameters for the \texttt{PLL} module (namely \texttt{PLLM} and \texttt{PPLN}), both described in Sec.~\ref{sec:clocking-schemes}.
Regarding the decoupling granularity, determining the most suitable value per layer depends on both board-related specifications (e.g., cache size) as well as code-related characteristics (e.g., number of output channels and kernel size).
In our case, we examine six different values, i.e., $ g\in\{0,2,4,8,12,16\}$, where $g=0$ indicates no DAE optimization and corresponds to the default input model.

Regarding the different clock alternatives, we define two parametric operating modes, namely the \textit{Low Frequency Operation} (LFO) and the \textit{High Frequency Operation} (HFO).
LFO exclusively employs the \texttt{HSE} clock source at a predefined frequency (50MHz) and aims to reduce power consumption on the board.
In contrast, HFO configures the system's clock using the \texttt{PLL} circuit, where the final \texttt{SYSCLK} value is determined by varying combinations of the \texttt{PLLN} and \texttt{PLLM} parameters, specifically $\texttt{PLLN}\in\{75,100,150,168,216,336,432\}$ and $\texttt{PLLM}\in\{25,50\}$.
This distinction between the two operating modes allows us to quickly transition between different \texttt{SYSCLK} frequencies, thus, mitigates susceptibility to the switching overhead associated with the \texttt{PLL}, as elaborated in Sec.~\ref{subsec:considerations}.
Overall, DVFS switching is performed between the memory (Lst.~\ref{dae-snippet-dw}:$3$) and compute (Lst.~\ref{dae-snippet-dw}:$7$) bound regions
, in order to exploit the decoupled access-execution transformation optimally, with LFO applied to the memory-bound and HFO to the compute-bound subsegments respectively. A similar approach is followed in pointwise convolution layers.

\textbf{DSE Insights:} Figure~\ref{fig:gran} shows the impact of varying operating frequencies (left) and granularity factors $g$ (right) on the layer latency and power consumption. 
First, we observe that as the number of operating frequency increases, the power consumption is traded for better performance, thus composing the design space. 
Moreover, changing values of the granularity factor can provide significant variation on the latency and the power consumption. For instance, power consumption can drop up to 54.2\% compared to the initial execution. Thus, the effective co-exploration of granularity $g$ and frequency leads to power/latency trade-offs.
The result of the DSE is a solution space per layer, where each solution trade-offs between performance and power consumption (\circled{2B}).
In this space, we select the Pareto optimal points, to be propagated to Step 3.

\subsection{Step 3: QoS-aware Energy Optimization}
\label{subsec:Optimizer}
In the final stage, we determine the optimal frequencies for each layer within the CNN, aiming to minimize the model's total energy consumption while satisfying a predefined latency budget (QoS).
We denote the set of all possible frequencies generated either through the \texttt{PLL} or the \texttt{HSE} clock as $F$ and the set of all possible granularity factors as $G$.
Let $n$ be the total number of layers of the CNN model and
$P_k=\{...,p^k_j=\{t^k_{j},E^k_{j}\},...\}, k\in\{1,...,n\}, j\in\{1,...,|P_k|\}$
be the set of Pareto optimal solutions (from Step 2) of layer $k$, where $t^k_{j}$ and $E^k_{j}$ denote the latency and the energy consumption of the $j^{th}$ pareto optimal solution for layer $k$ when operating with DVFS enabled with an HFO frequency $f\in F$ and granularity $g\in G$.
We consider the minimization of the overall energy $E$ of the CNN deployed on the target STM32 MCU, so that the overall execution time $T$ does not go beyond a user-defined $QoS$.
Then, the target optimization problem can be formulated as follows:
\vspace{-5pt}
\begin{flalign}
    \text{minimize} \quad & E=\sum_{k=1}^{n}\sum_{j\in P_k}E^k_jx_{kj} \\
    \text{s.t.} \quad & T=\sum_{k=1}^{n}\sum_{j\in P_k}t^k_jx_{kj} \leq QoS \\
    \quad & \sum_{j\in P_k}x_{kj}=1, \quad k=1,...,n
    \\
    \quad & x_{kj} \in {0,1}, \quad\quad k= 1,...,n, j\in P_k.
\end{flalign}

We model our problem according to the Multiple-Choice Knapsack Problem (MCKP)~\cite{kellerer2004multiple}, which extends the classical knapsack problem by categorizing items into distinct classes(\circled{3A}). 
In this formulation, the binary decision of including an item is replaced by the selection of precisely one item from each class and the goal is to maximize the value of items included in the knapsack while not exceeding its size.
In our case, each individual class represents the various Pareto optimal solutions $p^k_j$ per layer $k$. 
Each item in the class is characterized by its own value (i.e., energy consumption $E^k_j$) and size (i.e., latency $t^k_j$). 
Our goal is to minimize the overall energy consumption ($E$) while adhering to the size constraint ($T<QoS$).
We convert our minimization objective to a maximization one using the transformation found in~\cite{kellerer2004multiple}.
Last, we solve the optimization problem using a pseudo-polynomial time solution based on a
dynamic programming (DP) approach.
\section{Experimental Setup and Evaluation}
\label{sec:experimental}


\begin{figure}[b]
    \centering
    \includegraphics[width=\columnwidth]{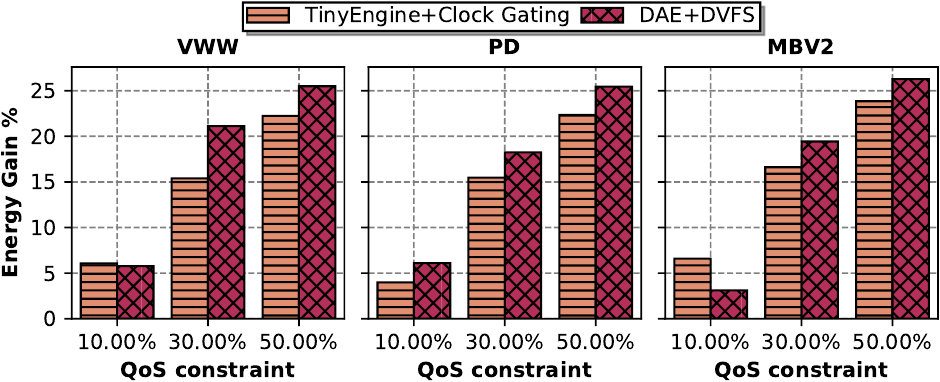}
    \caption{Energy consumption gains of our approach over the TinyEngine~\cite{lin2020mcunet} baseline. We compare against TinyEngine with Clock Gating over the examined CNN models.}
    \label{fig:comp_tiny_engine}
\end{figure}

\begin{figure*}[t]
    \centering
    \includegraphics[width=0.25\textwidth]{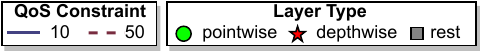}
    \vspace{-2pt}
    \\
    \centering
    \includegraphics[width=\textwidth]{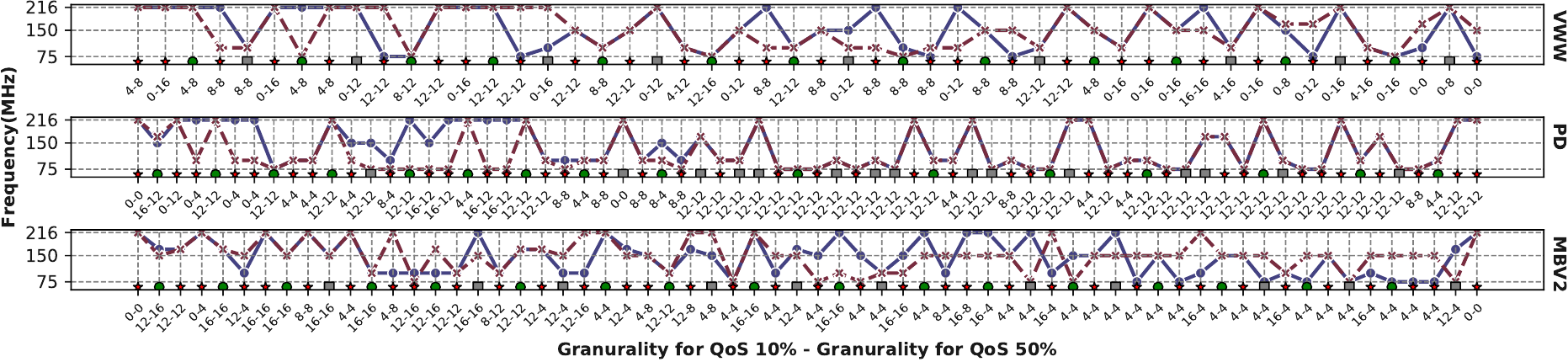}
    \caption{Frequency distribution throughout layer progression over the examined CNN models for 10\% and 50\% QoS constraints.}
    \label{fig:freq_over_time}
\end{figure*}

\textbf{Experimental Setup:} Our experimental evaluation is conducted on an STM32F767ZI Nucleo board, equipped with an ARM Cortex M7 CPU featuring a 16KB L1-cache. The board incorporates a High-Speed External (\texttt{HSE}) clock, ranging from 1MHz to 50MHz. For power consumption monitoring, we employed the INA219 power sensor. To mitigate potential variations arising from temperature-induced power fluctuations, we systematically compared each power measurement with the power consumption of the baseline input model at the corresponding timestamp. 
Our proposed methodology is evaluated over three pre-trained CNN models, namely Visual Wake Words (\texttt{VWW}), Person Detection (\texttt{PD}), and Mobilenet-V2 (\texttt{MBV2}), derived from the MCUNet~\cite{lin2020mcunet} inference library. Hardware-Aware Neural Architecture Search and linear int8 quantization have been used for quantization.
We conduct a comparative analysis between our approach and the state-of-the-art \texttt{TinyEngine}~\cite{lin2020mcunet}, which serves as the baseline for evaluation.
Our experiments are conducted in an iso-latency execution scenario, where we measure energy consumption for a certain period specified by a QoS constraint.
In the case of TinyEngine, this entails the board remaining in an idle state with a constant frequency of 216MHz after an inference, until the QoS threshold is met.
We also consider \texttt{TinyEngine} enhanced with clock gating, a technique designed to optimize power consumption by selectively deactivating non-utilized board clocks and the voltage regulator, thus minimizing power leakage throughout the CNN inference.

\textbf{Energy Comparison and Analysis:} Figure~\ref{fig:comp_tiny_engine} illustrates a comparison of energy consumption between our proposed approach, and the two configurations of the \texttt{TinyEngine}: one without any optimization and the other with clock gating applied. This evaluation encompasses various CNN inference models, each subject to discrete Quality of Service (QoS) constraints set at 10\% (tight), 30\% (moderate), and 50\% (relaxed). X-axis illustrates the QoS constraints, while Y-axis represents the normalized energy consumption. Our proposed approach surpasses both instances of the \texttt{TinyEngine}, exhibiting a reduction in energy consumption up to 25.2\%. Additionally, compared to the \texttt{TinyEngine} equipped with clock gating we achieve up to 7.2\% less energy consumption. Furthermore, our observations indicate that relaxing the QoS constraints can lead to a notable reduction in energy consumption, albeit at the cost of some performance trade-offs. For instance, when examining the Mobilenet-V2 model, the energy consumption of our approach under a relaxed 50\% QoS constraint decreases to 20.4\% compared to the stringent 10\% constraint.

\textbf{Frequency Scaling Analysis:} Figure~\ref{fig:freq_over_time} illustrates the HFO for each examined CNN. X axis indicates the corresponding layer type as the CNN execution proceeds and the granularities selected for 10\% and 50\% QoS constraint, respectively, while Y axis shows the operating frequency per layer. The LFO configuration at 50MHz of the memory-bound segment of each layer is excluded for simplicity. The observations derived are the following. Firstly, the operational frequency is configured to maximum (216MHz) mostly for performing pointwise convolutions, i.e. 58.8\% against 21.4\% for depthwise convolutions. The latter are less compute-intensive, thus decreasing the operational frequency will not lead to significant performance degradation. Furthermore, the 46.1\% of the pointwise convolutions and 43.4\% of the depthwise convolutions are executed over the lowest operating frequencies, i.e. 75MHz and 100MHz, aiming to boost the optimization objective of power minimization. Last but not least, we investigate the impact of QoS constraints on the operating frequency. Our experiments indicate that 18.6\% more layers are operating at 216MHz for tight constraints (10\%). Regarding the granularity analysis, for the relaxed QoS(50\%), 22.3\% more layers operate with granularity factor 16, compared to 10\% constraint. This is due to the fact that there is higher space to trade latency, thus the computation-bound parts are split to bigger segments, aiming to minimize switching overhead and provide power reduction.


    


\section{Conclusion}
\label{sec:conclusion}

In this work, we present a novel end-to-end methodology that exploits DVFS for optimizing the energy consumption of CNN inference on low-end STM32 MCUs. Our approach strongly leverages Decoupled Access Execute techniques to discretize memory-bound and compute-bound layer segments. Our approach outperforms existing state-of-the-art approaches by achieving up to 25.2\% less energy consumption.


\bibliography{sigproc} 

\begin{thebibliography}{10}

\bibitem{li2018edge}
E.~Li, Z.~Zhou, and X.~Chen, ``Edge intelligence: On-demand deep learning model co-inference with device-edge synergy,'' in {\em Proceedings of the 2018 Workshop on Mobile Edge Communications}, pp.~31--36, 2018.

\bibitem{ananthanarayanan2017real}
G.~Ananthanarayanan, P.~Bahl, P.~Bod{\'\i}k, K.~Chintalapudi, M.~Philipose, L.~Ravindranath, and S.~Sinha, ``Real-time video analytics: The killer app for edge computing,'' {\em computer}, vol.~50, no.~10, pp.~58--67, 2017.

\bibitem{howard2017mobilenets}
A.~G. Howard, M.~Zhu, B.~Chen, D.~Kalenichenko, W.~Wang, T.~Weyand, M.~Andreetto, and H.~Adam, ``Mobilenets: Efficient convolutional neural networks for mobile vision applications,'' {\em arXiv preprint}, 2017.

\bibitem{amazonsagemakeredge}
``Amazon sagemaker edge.'' \url{https://aws.amazon.com/sagemaker/edge/}.

\bibitem{googleedgetpu}
``Google edge tpu.'' \url{https://cloud.google.com/edge-tpu}.

\bibitem{ZigZagMeiHJGV21}
L.~Mei, P.~Houshmand, V.~Jain, S.~Giraldo, and M.~Verhelst, ``Zigzag: Enlarging joint architecture-mapping design space exploration for dnn accelerators,'' {\em IEEE Transactions on Computers}, 2021.

\bibitem{Kwon2020MAESTRO}
H.~Kwon, P.~Chatarasi, V.~Sarkar, T.~Krishna, M.~Pellauer, and A.~Parashar, ``Maestro: A data-centric approach to understand reuse, performance, and hardware cost of dnn mappings,'' {\em IEEE micro}, 2020.

\bibitem{banbury2021micronets}
C.~Banbury, C.~Zhou, I.~Fedorov, R.~Matas, U.~Thakker, D.~Gope, V.~Janapa~Reddi, M.~Mattina, and P.~Whatmough, ``Micronets: Neural network architectures for deploying tinyml applications on commodity microcontrollers,'' {\em Proceedings of Machine Learning and Systems}, 2021.

\bibitem{kakolyris2023road}
A.~K. Kakolyris, M.~Katsaragakis, D.~Masouros, and D.~Soudris, ``Road-runner: Collaborative dnn partitioning and offloading on heterogeneous edge systems,'' in {\em 2023 Design, Automation \& Test in Europe Conference \& Exhibition (DATE)}, pp.~1--6, IEEE, 2023.

\bibitem{chitty2022neural}
K.~T. Chitty-Venkata and A.~K. Somani, ``Neural architecture search survey: A hardware perspective,'' {\em ACM Computing Surveys}, 2022.

\bibitem{david2021tensorflow}
R.~David, J.~Duke, A.~Jain, V.~Janapa~Reddi, N.~Jeffries, J.~Li, N.~Kreeger, I.~Nappier, M.~Natraj, T.~Wang, {\em et~al.}, ``Tensorflow lite micro: Embedded machine learning for tinyml systems,'' {\em Proceedings of Machine Learning and Systems}, vol.~3, pp.~800--811, 2021.

\bibitem{microsoftnni}
``Microsoft nni.'' \url{https://github.com/microsoft/nni}.

\bibitem{lai2018cmsis}
L.~Lai, N.~Suda, and V.~Chandra, ``Cmsis-nn: Efficient neural network kernels for arm cortex-m cpus,'' {\em arXiv preprint arXiv:1801.06601}, 2018.

\bibitem{capotondi2020cmix}
A.~Capotondi, M.~Rusci, M.~Fariselli, and L.~Benini, ``Cmix-nn: Mixed low-precision cnn library for memory-constrained edge devices,'' {\em IEEE Transactions on Circuits and Systems II: Express Briefs}, 2020.

\bibitem{de2022depth}
J.~D. De~Leon and R.~Atienza, ``Depth pruning with auxiliary networks for tinyml,'' in {\em ICASSP 2022-2022 IEEE International Conference on Acoustics, Speech and Signal Processing (ICASSP)}, IEEE, 2022.

\bibitem{moosmann2023flexible}
J.~Moosmann, H.~Mueller, N.~Zimmerman, G.~Rutishauser, L.~Benini, and M.~Magno, ``Flexible and fully quantized ultra-lightweight tinyissimoyolo for ultra-low-power edge systems,'' {\em arXiv preprint:2307.05999}, 2023.

\bibitem{spantidi2023automated}
O.~Spantidi and I.~Anagnostopoulos, ``Automated energy-efficient dnn compression under fine-grain accuracy constraints,'' in {\em 2023 Design, Automation \& Test in Europe Conference \& Exhibition (DATE)}, 2023.

\bibitem{saha2020rnnpool}
O.~Saha, A.~Kusupati, H.~V. Simhadri, M.~Varma, and P.~Jain, ``Rnnpool: Efficient non-linear pooling for ram constrained inference,'' {\em Advances in Neural Information Processing Systems}, vol.~33, pp.~20473--20484, 2020.

\bibitem{liberis2021munas}
E.~Liberis, {\L}.~Dudziak, and N.~D. Lane, ``$\mu$nas: Constrained neural architecture search for microcontrollers,'' in {\em Proceedings of the 1st Workshop on Machine Learning and Systems}, pp.~70--79, 2021.

\bibitem{lin2020mcunet}
J.~Lin, W.-M. Chen, Y.~Lin, C.~Gan, S.~Han, {\em et~al.}, ``Mcunet: Tiny deep learning on iot devices,'' {\em Advances in Neural Information Processing Systems}, vol.~33, pp.~11711--11722, 2020.

\bibitem{jaiswal2023minun}
S.~Jaiswal, R.~K.~K. Goli, A.~Kumar, V.~Seshadri, and R.~Sharma, ``Minun: Accurate ml inference on microcontrollers,'' in {\em Proceedings of the 24th ACM SIGPLAN/SIGBED International Conference on Languages, Compilers, and Tools for Embedded Systems}, pp.~26--39, 2023.

\bibitem{zhao2018deepthings}
Z.~Zhao, K.~M. Barijough, and A.~Gerstlauer, ``Deepthings: Distributed adaptive deep learning inference on resource-constrained iot edge clusters,'' {\em IEEE Transactions on Computer-Aided Design of Integrated Circuits and Systems}, vol.~37, no.~11, pp.~2348--2359, 2018.

\bibitem{snowdon122009platform}
D.~C. Snowdon12, E.~Le~Sueur, S.~M. Petters, and G.~Heiser, ``A platform for os-level power management,'' {\em The European Professional Society on Computer Systems 2009}, 2009.

\bibitem{acun2019fine}
B.~Acun, K.~Chandrasekar, and L.~V. Kale, ``Fine-grained energy efficiency using per-core dvfs with an adaptive runtime system,'' in {\em 2019 Tenth International Green and Sustainable Computing Conference (IGSC)}, pp.~1--8, IEEE, 2019.

\bibitem{jejurikar2004leakage}
R.~Jejurikar, C.~Pereira, and R.~Gupta, ``Leakage aware dynamic voltage scaling for real-time embedded systems,'' in {\em Proceedings of the 41st annual Design Automation Conference}, pp.~275--280, 2004.

\bibitem{gay2018beginning}
W.~Gay, ``Beginning stm32,'' {\em Beginning STM32}, 2018.

\bibitem{kellerer2004multiple}
H.~Kellerer, U.~Pferschy, D.~Pisinger, H.~Kellerer, U.~Pferschy, and D.~Pisinger, ``The multiple-choice knapsack problem,'' {\em Knapsack Problems}, pp.~317--347, 2004.

\end{thebibliography}
\bibliographystyle{ieeetr}

\end{document}